\documentclass{aa501}
\usepackage{epsfig}

\newcommand{\Msun}{M_\odot}
\newcommand{\be}{\begin{equation}}
\newcommand{\ee}{\end{equation}}
\newcommand{\bea}{\begin{eqnarray}}
\newcommand{\eea}{\end{eqnarray}}

\begin{document}

\title{
Non-Friedmann cosmology
for the Local Universe, \\ significance of the universal Hubble
constant\\ and short-distance indicators of dark energy}
\subtitle{}
\author{
A.D. Chernin \inst{1,2,3},
P. Teerikorpi \inst{2},
and Yu.V. Baryshev \inst{4,5}}

\institute{Sternberg Astronomical Institute, Moscow University, 119899,
Moscow, Russia
\and
Tuorla Observatory,
University of Turku, FIN-21500 Piikki\"o, Finland
\and
Division of Astronomy, University of Oulu, FIN-90014, Finland
\and
Institute of Astronomy, St.Petersburg State University,
Staryj Peterhoff, 198504, St.Petersburg, Russia
\and
Isaac Newton Institute of Chile, Saint-Petersburg Branch, Russia
}

\date{Received / Accepted}

\abstract{Basing on the increasing evidence for the cosmological
relevance of the local Hubble flow, we consider a simple analytical
cosmological model for the Local Universe. This is a non-Friedmann
model with a non-uniform static space-time. The major dynamical
factor controlling the local expansion is the
antigravity produced by the omnipresent and permanent dark
energy of the cosmic vacuum (or the cosmological constant). The
antigravity dominates at distances larger than 1-2 Mpc from the
center of the Local group. The model gives a natural
explanation of the two key quantitative characteristics of the
local expansion flow, which are the local Hubble constant and the
velocity dispersion of the flow. The observed kinematical
similarity of the local and global flows of expansion is
clarified by the model. We demonstrate analytically the
efficiency of the vacuum cooling mechanism
that allows one to see the Hubble law so close to the Local group.
Special
significance is argued for the `universal Hubble constant' $H_V$, depending
only on the vacuum density ($H_V \simeq 60 $
km/s/Mpc). The model makes a number of testable predictions.
It also shows unexpectedly that it is
the dwarf galaxies of the local flow with the shortest distances
and lowest redshifts that may be the most sensitive indicators of
dark energy in our neighborhood.
}

\titlerunning{Non-Friedmann cosmology for the Local Universe}

\maketitle

\section{Introduction}

Cosmology has changed dramatically with the discovery of the
cosmic acceleration and dark energy of the cosmic vacuum
(or the cosmological constant) from observations of high-redshift
supernovae (Riess et al. 1998; Perlmutter et al. 1999).
A new `concordance cosmological model' has emerged which describes the
accelerating cosmological expansion on cosmological scales
beyond the cell of uniformity ($> 300-1000$ Mpc).
The model is in good agreement with all
cosmological data including the WMAP results on the CMB
anisotropy (Spergel et al. 2003). The Friedmann theory with a
non-zero positive cosmological
constant serves as a reliable basis for the concordance
model, the present-day standard model in cosmology (see
e.g. reviews by Chernin 2001; Peebles \& Ratra
2003).

A new field of studies in which the cosmic vacuum has
proved to be important, lies deep inside the cell of
uniformity. In this Local Volume (distances $< 10-20$ Mpc)
cosmological expansion was originally discovered. Our
understanding of the physical conditions there has improved so
much that, for the first time, we have
reliable grounds for finding a solution to the
Hubble-Sandage paradox. The paradox is in the fact that
cosmological expansion with a regular linear velocity-distance
relation was discovered by Edwin Hubble where it should not be. Indeed, the
Friedmann theory predicts and describes the expansion of a
smooth, homogeneous self-gravitating medium. But locally,
the matter is far from uniform and rather reveals a
fairly chaotic spatial distribution of galaxies. Therefore the
Friedmann theory cannot be applied to the Local Volume. The paradox
was splendidly realized by Allan Sandage (Sandage et al. 1972, Sandage
1986, 1999). He saw a most striking mystery in the fact that the
local expansion proceeds in a quite regular way: the linear
velocity-distance relation (the Hubble law) with a low
velocity dispersion is clearly observed here. Moreover, the local
rate of expansion is similar, if not identical, to the
global rate characterized by the Hubble constant.

Recently, Thim et al. (2003) have emphasized a related problem:
there is no evidence for the expected turn-down of the velocity-distance
relation, when one approaches the Local Group (c.f. Ekholm et al.
2001). Whiting (2005) notes
the similar absence of infall flow around local galaxy groups in general.

Soon after the discovery of the cosmic vacuum, we suggested
(Chernin, Teerikorpi, Baryshev 2000 -- hereafter Paper I) that
the cosmic vacuum with its omnipresent and perfectly uniform
density might provide the dynamical background for a regular quiescent
expansion in the Local Volume. Our key argument
comes from the fact that the vacuum antigravity dominates over
the matter gravity of the Local Group at the distances larger than
1-2 Mpc from the center of the group (Chernin 2001, Baryshev,
Chernin, Teerikorpi 2001, Karachentsev, Chernin, Teerikorpi
2003). The increasing observational evidence and theoretical
arguments for this view have been recently discussed by
Teerikorpi, Chernin, Baryshev (2005).
 In that paper it was also shown that dark energy may explain still
another problem in the Local Volume: why peculiar velocities
do not depend on luminosity (mass) (Karachentsev \& Makarov 1996;
Whiting 2003; 2005). 

This new approach has extended cosmology
into the deep interior of the cosmic cell of uniformity.
Especially, the Local Volume must be regarded as the nearest
cosmologically relevant object. So we find it good to use
the term `the Local Universe' for our relatively small `Local Volume'
sample of the Universe which, however, is
fairly typical for galaxy space in general. We will also
speak about `local cosmology',
assuming at the same time that the Universe as a whole is described by the
concordance model of the traditional `global' cosmology.

In the present paper, we develop our approach further and study
a simple analytical non-Friedmann model for the
Local Universe (Secs.2,3). The model naturally explains
two key characteristics of the local expansion
flow: the local Hubble constant and the velocity
dispersion (Sec.4). A special significance is argued for the universal
'vacuum' Hubble constant (Sec.5). The model predicts
that the cosmic antigravity and
vacuum density can be detected and measured in
our nearest neighbourhood (Sec.6). In Sec.7,
complementary computer simulations are briefly reviewed and a
general discussion of the results is given. We summarize the conclusions
are in Sec.8.

\section{Local cosmology: Newtonian theory }

It is clear from the considerations of Sec.1 that a non-Friedmann
counter-part of Friedmann's cosmological model is needed for the
Local Universe where the matter distribution is highly
non-uniform. In constructing a non-Friedmann
theory, we focus first on the nearest and most
important distances, $1-5$ Mpc, where the observational data are
fairly complete and distance determinations are accurate enough
(Karachentsev \& Makarov 2001,
Karachentsev et al. 2002, 2003a, Thim et al. 2003, 
Karachentsev et al. 2004; Paturel \&
Teerikorpi 2005, Rekola et al. 2005). We start with a simple
version of the theory in which the following physical conditions
are assumed for the present epoch of cosmic evolution:

1. The environment of the Local Universe (i.e. the distribution
and motion of matter outside it) does not affect essentially the
dynamics inside it and one may neglect the matter
distribution and motions at the distances larger than, say, 5 Mpc,
in the first and main approximation.

2. The matter mass of the Local Universe is collected  almost
entirely -- with a few percent accuracy -- within the area of
the Local Group at the distances $\le 1$ Mpc from the barycenter
of the group, in this approximation.

3. The gravity field produced by the Local Group is
spherically symmetrical and static at the distances $\ge 1.5-2$
Mpc from the group barycenter (more about this below).

4. The interaction of the out-flowing dwarf galaxies with each other is
negligible.

5. The flow consists of the galaxies which have escaped from the
gravitational potential well of the Local Group.

\subsection{The equation of motion}

Under these simplified, but for the present discussion
sufficiently reasonable assumptions, the dynamics at
distances 1-5 Mpc from us is reduced to the problem of a test
particle motion in the gravity field of the Local Group treated
as a point-like mass on the background of the antigraviting
vacuum. Gravity of the Local Group produces the radial force $-G
M/r^2$, where $r$ is the distance from the group barycenter and
$M = 1.5 \times 10^{12} \Msun$ is the group mass. The antigravity
of vacuum produces the radial force $G 2\rho_V (\frac{4
\pi}{3}r^3)/r^2$, where $-2 \rho_V = \rho_V + 3 p_V$ is the
effective gravitating density of vacuum, according to General
Relativity  (see e.g. Chernin 2001), and $\rho_V = 7
\times 10^{-30}$ g/cm$^3$ is the concordance value of
the vacuum density.

The radial
component of motion in this gravity/antigravity force field
obeys the Newtonian equation:
\be \ddot r(t, \chi) = - GM/r^2 + r/A^2, \ee where $r(t, \chi)$
is the distance of a particle to the barycenter of the Local
group, $\chi$ is the Lagrangian coordinate of the particle; the
constant
\be A = (\frac{8 \pi G}{3} \rho_V)^{-1/2} \simeq 5 \times 10^{17} \;\;s
\simeq 1.5 \times 10^{28} \;\; cm \ee
is the characteristic vacuum time/length; hereafter the speed of
light $c = 1$ is used in the formulas. It is interesting that the
constant vacuum time/length is numerically near the current
cosmic age or the distance to the cosmological horizon: $t_0 \sim
A$; a discussion of this fact may be found, for instance, in
Chernin (2005).

It is seen from Eq.1 that the gravity force ($\propto 1/r^2$)
dominates over the antigravity force ($\propto r$) at smaller
distances; the acceleration is negative there. At the distance
\be R_V = (GM A^2)^{1/3} = (\frac{3}{8\pi}M/\rho_V)^{1/3} \simeq 1.3 \;\;
Mpc \ee the gravity and antigravity balance each other, so the
acceleration is zero at the `zero-gravity sphere' of the radius
$R_V$ (Paper I). At larger distances, $r > R_V $, antigravity
dominates, and the acceleration is positive.

We will study mostly the region outside the zero-gravity sphere
where the gravity field is assumed to be static and spherically
symmetrical (see above). In fact, computer simulations of
the Local group and its gravity (Dolgachev et al. 2003, Chernin
et al. 2004) show that the zero-gravity surface is fairly
spherically symmetrical at the present epoch and it is
approximately (with a 10-15\% accuracy, at least) unchanged in
size and shape over more than 12 Gyr of the history of the Local
Group.

The first integral of Eq.1 expresses, as usual, the mechanical
energy conservation:
\be \frac{1}{2}\dot r^2 = GM/r + \frac{1}{2}(r/A)^2 + \bar E, \ee
where $\bar E(\chi)$ is the total mechanical energy of a particle
with the Lagrangian coordinate $\chi$ (per its unit mass).
Here the potential energy
\be U(r) = - GM/r -\frac{1}{2}(r/A)^2 \ee
is negative, and both matter and vacuum give to $U$
contributions of the same sign (contrary to the right-hand side
of Eq.1). Because of the vacuum, the gravitational potential cannot be
normalized to $U = 0$ at $R = \infty$, and $U(R)$ goes to
$-\infty$ in both limits of $R \rightarrow 0$ and $R \rightarrow
\infty$ which is not a standard situation in mechanics. It means,
in particular, that the trajectories with $E < 0$ are not
necessarily finite. Such a behaviour of the potential has a
clear analogue in General Relativity applied to the same
problem (sect.3).

\subsection{Radial trajectories}

The dwarf galaxies of the local expansion flow move along
approximately radial trajectories in the potential $U(r)$. The
total energy of each particle that has escaped from the gravity potential well
of the Local Group exceeds the maximal value of the potential $U$:
\be E > U_{max} = -\frac{3}{2} GM/R_V.  \ee
Asymptotically, when $r$ goes to infinity, and the vacuum
dominates entirely, the solution of Eq.4 is exponential,
\be r (t,\chi) \propto \exp{t/A}, \ee
so that the linear velocity-distance law, $\dot r = r/A_V$,
appears with the constant expansion rate:
\be H_V \equiv 1/A \simeq 60 \;\;\; \mathrm{km/s/Mpc}. \ee

The quantity $H_V$ is a universal physical constant
directly related to the vacuum density (or the cosmological
constant); its provisional numerical value, here adopted,
comes from the concordance data on $\rho_V$. We call $H_V$
the universal Hubble constant.

Generally, integrating Eq.4, one gets:
\be t-T(\chi) = \int{[2GM/r + (r/A)^2 + 2\bar E]^{-1/2} d r}, \ee
where $T(\chi)$ is the second integration constant. Two
time-independent functions of the Lagrangian coordinates, $\bar
E(\chi)$ and $t_0(\chi)$, are determined by the particle's "initial
conditions" that are its velocity and distance from the center
at a given moment of time.

In the simplest case of the parabolic ($\bar E = 0$) motion, the exact
integral of Eq.9 is expressed in terms of elementary functions:
\be r (t, \chi) \propto [sinh (3t/2A)]^{2/3}. \ee

\subsection{Local vs. global}

It is significant that the mathematical structure of Eq.4 is
similar to that of the Friedmann cosmological equation for the
global Friedmann radius $R(t)$:
\be \frac{1}{2}\dot R (t, \chi)^2 = G M(\chi)/R + \frac{1}{2}
R^2/A^2 +  E(\chi), \ee where $M(\chi)$ is the
time-independent mass of the non-relativistic matter within the
sphere of the radius $R(\chi,t)$, and $\chi$ is the Lagrangian
coordinate which corresponds to the Eulerian coordinate
$R(t,\chi)$; the time-independent value $E(\chi) = k (R/a)^2$,
$a$ is the curvature radius, $-k = -1,0,+1$ is the sign of the 3D
space curvature. As is well-known, the function $R(t, \chi)$ is a
product of the curvature radius $a(t)$ and a function of the
Lagrangian coordinate only.
In terms of physics, the difference of Eq.11 from Eq.4 is in
the dependence of the mass
$M (\chi)$ on the Lagrangian coordinate in the Friedmann equation,
due to the uniform matter distribution.
According to the non-Friedmann Eq.4, each body moves in the
potential of the same gravitational mass.
 Clearly, the solution of Eq.11 has the same -- in mathematical
structure -- form as Eq.10 for the parabolic
($E = 0$) motion.

It is especially suggestive that, in the limit of large
$R$ and large $r$, both equations lead to the same exponential
solution with the linear velocity-distance relation and the
universal Hubble constant $H_V = 1/A$. This similarity
demonstrates that both local and global expansion flows are
controlled by the same physical factor -- the cosmic vacuum with
its perfectly uniform density, -- when the vacuum dominates
entirely. In this limit, the global cosmological expansion and the
local Hubble flow are parabolic. The only constant physical
parameter for the both is the universal Hubble constant, in
this asymptotic regime.

It is also remarkable that the real Local Universe and the global
Universe are both now rather close to this common limit.
Thanks to this underlying physical cause the
observed values of $H_{\mathrm{local}}$ and $H_0$ are so close to
each other: they are close because each of them is close
to $H_V$.

Equations 1-10 constitute the basic equations of our non-Friedmann
model for the Local Universe. To describe the present-day local flow,
we may put $t = t_0 \simeq 14$ Gyr in
Eqs.1-10. Then we have the dependence of the acceleration and
velocity of the flow particles on the distance $r$ at the
present epoch. Especially informative is the phase
(velocity-distance) relation given by Eqs.4,10, showing that
the velocity of a particle is determined by its distance
and energy $E$, while the distance itself depends on both energy
and the `initial moment' $T$ that are different for different
particles (e.g. particles emerge at different times from within
the zero-gravity surface). This set of equations in combination
with observational data gives a clear picture of the
evolution and the present kinematics of the expansion flow in
the gravity/antigravity force field of the Local Universe. They
clarify also the major trends in the evolution of the flow during
the whole life time ($\simeq 12-13$ Gyr) of the Local group (see
below).

\section{Local cosmology: General Relativity theory}

In the Newtonian theory for the local Universe (Sec.2), the dark
energy is treated as relativistic `fluid', in
accordance with the General Relativity, while the ordinary matter
and the space-time in which the matter move are described in
terms of non-relativistic physics. Now we give a completely
relativistic theory for the Local Universe.

\subsection{Spherically symmetric static metric}

It is most important that the set of the Newtonian spherically
symmetrical equations (Eqs.1-10 above) has an exact analogue in
General Relativity. Indeed, there is a well-known GR solution for
a point-like body on the vacuum background:
\be ds^2 = F(r)dt^2 - r^2 d\Omega^2 - F(r)^{-1} dr^2, \ee where
\be F(r) = 1 - 2GM/r - (r/A)^2. \ee This time-independent
metric describes a spherically-symmetrical static
space-time. Identifying here the constant parameters $M$ and $A$
with the mass of the Local Group and the vacuum time/length,
respectively, we may use this metric for the local cosmology. The
local flow particles move along geodesics in the space-time of
Eqs.12,13.

In the limit of small deviations from the Newtonian (Galilean)
space-time (quite appropriate for the Local Universe),
the metric Eq.13 takes the form:
\be F(r) \simeq 1 - GM/r - \frac{1}{2} (r/A)^2 = 1 + U(r), \ee
where $U(r)$ is the gravitational potential of the Newtonian
theory and exactly the same as in Eq.5.

The metric of Eqs.12,13 representing the static GR model of the
Local Universe contains all information about possible
motions in this static space-time. In particular, the radial
motions of the local flow particles are naturally the same
(practically exactly) as in the Newtonian version (Eqs.1-10) of
the model.

\subsection{Embedding the Local Universe into global space}

Furthermore, it is important that
the 3D volume of the Local Universe with the metric of Eqs.12-14
may be considered as `embedded' in the Friedmann global 3D
cosmological space with the non-static metric
\be ds^2 = dt^2 - a(t)[d\Omega^2 + d\chi^2], \ee where $a(t)$ is
the 3D curvature radius and/or the expansion scale factor (for $k
= 1$) which is proportional to the Friedmann radius $R(t, \chi)$ of Eq.11.

Indeed, the Local Universe may be described as a spherical
`vacuole' in the uniform distribution of matter, as given by the
exact Einstein-Straus (1945) solution. The size of the vacuole
$R_L(t, M)$ increases with time in accordance with the Friedmann
expansion solution $R_L \propto a(t)$, and the central mass
$M$ is equal to the mass of the (non-relativistic) matter within
the sphere of radius $R_L$ in the smooth matter distribution
of the Friedmann Universe:
\be M = \frac{4\pi}{3} \rho(t)R_L(t)^3 = Const.\ee

In the vacuole model with the conjunction condition of Eq.16, the
environment of the Local Universe does not affect the dynamics
inside it (c.f. condition 1 assumed in Sec.2). The
Einstein-Straus (1945) solution demonstrates how the local
non-Friedmann static cosmological model
may be compatible with the Friedmann metrics (and dynamics) of
the global cosmological expansion.

 It is easy to see that the zero-gravity surface has the radius
 $R_V =
\frac{1}{2}(\rho_M/\rho_V)^{1/3} R_L(t_0) \simeq
(\frac{3}{14})^{1/3} R_L(t_0)$, where $R_L(t_0)$ is the vacuole
radius at present. The distance interval $(0.6 - 1)R_L(t_0)$
is dominated by the vacuum in the Local
Universe.

Arguing further along this line, one can imagine that the whole
Universe may contain not one, but many (or infinite
number) of vacuoles of various sizes and
masses. Moreover, a picture is theoretically possible in which
vacuoles fill almost all cosmic space without intersecting
each others. Such a complex non-uniform expanding structure is
exactly(!) described by the equations above. Obviously
in this picture the tiny contribution of uniform matter distribution
between vacuoles can be neglected, and
one obtains a highly non-uniform, but completely regular,
cosmological model. The model describes the global
expansion in terms of the relative motions of discrete masses
which are in the centres of the vacuoles. The masses move apart
from each others on the uniform vacuum background, in
accordance with the Hubble law and the expansion factor given by
the Friedmann theory (Eq.11). In fact, because the zero-gravity
radius is inside the vacuole in such a model, each vacuole
expulses every other one and the expansion is generally
accelerating (c.f. sect.7.2 in Teerikorpi, Chernin, Baryshev
2005).

In this idealized model, the global Hubble constant,
$H = \dot R/R = \dot a/a$, is exactly the same for any two masses of
the global expansion flow. In accordance with the Friedmann
theory (Eq.7), it is presently ($t=t_0$):
\begin{eqnarray}
 H(t_0) \equiv H_0 &=& H_V [1 + \rho_M (t_0)/\rho_V]^{1/2} \simeq
H_V (1 + 3/7)^{1/2} \nonumber\\
&\simeq& 1.2 H_V \simeq 72 \;\; \; km/s/Mpc,
\end{eqnarray}
in good agreement (not surprisingly) with the
concordance data (Spergel et al. 2002); it is also taken
into account that $E = 0$ in Eq.11, according to the same
data ($\rho_M = 3 \times 10^{-30}$ gcm$^{-3}$ is the
concordance value for the mean matter (dark matter + baryons)
density at present).

In terms of General Relativity, the asymptotic similarity of the
local and global expansion flows is seen as clearly as in the
Newtonian treatment of Sec.3. Indeed, the static metric of
Eqs.12,13 and the expanding metric of Eq.15 have a common
asymptotic in the limit when and where the vacuum antigravity
dominates entirely. This limit is described by de
Sitter's static solution. As is well-known, this solution has the
metric of Eq.12 with $F(r)= 1 - (r/A)^2$. The space-time of de
Sitter's solution is determined by the vacuum alone which is ever
static itself, and the vacuum controls completely both space-time
metric and any geodesic in this space-time.

(Note that the vacuole model can be generalized: instead
of the central mass one may use a spherically symmetrical matter
distribution of the same total mass, as in the
Tolman-Bondi solution that is discussed e.g. by Gromov et al.
(2001) in the context of the Local Universe.)

\section{The efficiency of vacuum cooling}

As demonstrated in Secs.2,3, the expansion flow described by the
local cosmological model tends to a regular kinematical structure
with the linear velocity-distance relation, $\dot r = r/A$, and
the universal Hubble constant $H_V = 1/A$, at the limit of the
cosmic vacuum domination. It is important to emphasize that the
regular self-similar parabolic regime is a generic asymptotic for
all the expansion flow motions, independently of their initial
conditions, in the Local Universe.

\subsection{From initial chaos to asymptotic order}

Indeed, in the model of Sec.2,3, the initial conditions are
given in terms of the functions $\bar E(\chi)$ and
$T(\chi)$. It is obvious from Eqs.4,9 that asymptotically
both functions become negligible, since $E/r^2 \rightarrow 0$ and
$T/t \rightarrow 0$, in this limit. The self-similar motion
described by Eq.10 is a kind of dynamical attractor for a wide
variety of motions possible in the gravity/antigravity potential
of the Local Universe. If the energy condition of Eq.6 is met,
any particle escapes from the interior of the zero-gravity sphere
and its motion reaches asymptotically the parabolic regime with
the universal Hubble constant.
This basic evolutionary trend in the
local flow drives it from initial chaos to asymptotic order
(see also Chernin et al. 2004, 2005). It is also because this
internal trend
is so quick that the velocity dispersion of the local flow
is rather low.

To put this in a quantitative way, let us introduce a `random'
velocity $v$ as a difference between the velocity $\dot r$ and a
`regular' (asymptotic) velocity $r/A_V$. Then the velocity $v$ at
the zero-gravity distance $r = R_V$ is
\be v (R_V) = v_1 = (R_V/A)[(3 + \bar E A^2/R_V^2)^{1/2} -1]. \ee

When the trajectory of a body reaches a distance $r$ outside the
zero-gravity surface, the random velocity $v(r)$ (for an
arbitrary $\bar E$) is given by the relation:
\bea
 v (r) = &(R_V/A)[(r^2/R_V^2 + 2 + 2R_V/r + 2 v_1A/R_V +& \nonumber\\
         &v_1^2A^2/R_V^2)^{1/2} - r/R_V].&
\eea
Note that different values of the total energy $E$ give an
`initial' distribution of the velocity $v$.  A comparison of
$v(r)$ with $v_1$ enables us to see that the random velocity
decreases along the trajectory and approaches asymptotically zero
no matter what its initial value $v_1$ is.

\subsection{The vacuum cooling factor}

With the `vacuum cooling factor', $q_V \equiv v_1/v$, we
may give a measure to the efficiency of the cooling mechanism:
\bea
 q_V = &(v_1 A/R_V)[(r^2/R_V^2 +2 + 2R_V/r + 2 v_1A/R_V +& \nonumber\\
       &v_1^2A^2/R_V^2)^{1/2} - r/R_V]^{-1}.&
\eea
In the simplest case of the parabolic motion ($\bar E = 0$), the
velocity $v_1 = (\sqrt{3} -1) R_V/A_V$, and we have:
\be v (r) = (R_V/A)[(r^2/R_V^2 +2 R_V/r)^{1/2} - r/R_V]. \ee Then
the cooling factor in the parabolic expansion flow is
\be q_V (r) \equiv v_1/v(r) = (\sqrt{3} - 1)[(r^2/R_V^2 +2)^{1/2}
- r/R_V]^{-1}. \ee One may see that the random velocity is
diminished by factor 3 during the time when the body covers the
path from $r = R_V$ to the distance $r \simeq 2 R_V \simeq 3$
Mpc. When the same body reaches, say, distances 4 or 6 $R_V$, the
cooling factor increases to $q_V = 12$ and $q_V = 56$,
correspondingly.

Now we compare the vacuum cooling with the usual
cosmological adiabatic cooling. The latter is described by the
relation $v a(t) = $ constant, which gives only the factors 2, 4
and 6 when the scale factor $R(t)$ increases 2, 4 or 6 times;
this is considerably less than the corresponding factors 3, 12 and 56
in the expansion flow of our model.

It is also interesting to follow -- for a contrast -- the dynamics
of the same expansion, but in the absence of vacuum. With the
same statement of the problem as above, but with $\rho_V = 0$, we
would have:
\be \dot r^2 = 2GM/r + 2\bar E. \ee When, for instance,
$\bar E > 0$, the asymptotic (as $r$ goes to infinity) motion is
inertial and the velocity-distance law is a linear one: $\dot r =
r/t$, where the `Hubble factor', $H = 1/t$, depends on time only.
This means that the flow may cool down without the vacuum
just because of its expansion. But the efficiency is then much lower.
Indeed, taking $r/t$ as a regular
expansion velocity, we may introduce the random velocity $v$ by the
relation $\dot r = r/t + v$. Then
\be v = (2E)^{1/2} [(1 + \frac{GM}{2E r})^{1/2} -1]. \ee In the
first approximation for $v/(2E)^{1/2} < 1$, this leads to the
`adiabatic' relation $vr =$ constant. As a result, the cooling
factor in the absence of vacuum, $q \propto r$, is significantly
smaller than the vacuum cooling factor $q_V$ above.

Thus, vacuum cooling is very efficient and results in a
dynamical evolution of the outflow from chaos to order. When
applied to the nearest distances 1-3 Mpc, the
non-Friedmann cosmology (for parabolic expansion) gives a
typical velocity dispersion between $v_1 = (\sqrt{3}-1)
R_V/A = 60$ km/s at $r = R_V$ and $v(r=2 R_V) = \frac{1}{3} v_1 =
20$ km/s at $r = 2R_V$, which is near the observed figure 30-60 km/s
(Karachentsev \& Makarov 2001;
Karachentsev et al. 2002, 2003a; Ekholm et al. 2001;
Paturel \& Teerikorpi 2005).

\section{Significance of the universal Hubble constant}

In this and the next sections, we discuss new observational
prospects offered by the local cosmology. The background is
naturally the global cosmology where both $H_V$
and $H_0$ are among the basic parameters.
As we show here, there are good prospects for the uniform
vacuum density and the universal $H_V$ to be determined locally, but the
exact value of the Hubble constant $H_0$ refers to the average
global expansion rate that cannot be measured in a local,
inhomogeneous environment. Locally, a good advantage is the high, 10 \% or
better, accuracy of distance measurements, while on large scales
distances continue to be hampered by low accuracy and
systematic errors.

\subsection{The stochastic Hubble expansion rate}

Of course, a "precision cosmology" value of $H_0$ may be
obtained from the analysis of the CMBR.
However, this still depends on assumed cosmological physics and
dark matter and dark energy components (Spergel et al. 2003), and
it would be very important to have an independent direct measurement on
large scales. Deep space measurements of $H_0$, even bypassing
local calibrations, from the Sunuyaev-Zeldovich effect and the
time delay in gravitational lens images, are still affected by
rather large systematic errors. The same must be said about
the proposed way of studying the distance scale
via the Eddington efficiency of luminous quasars (Teerikorpi
2005).

In intermediate and closer space, the HST Key Project (Freedman et al.
2001) derived $H_0 = 72 \pm 8$ km/s/Mpc, using secondary
distance indicators calibrated from local galaxies whose distances
are known from Cepheids. This value agrees with the value
from the CMBR analysis and is a part
of the concordance picture. However, one should not lose from sight
the fact that $H_0$ has not yet been determined directly on large scales
with a high accuracy and there are still possible sources of systematic
error in the HST value for $H_0$ (e.g. Paturel \& Teerikorpi 2005).

Further we wish to emphasize that
the measured Hubble constant is not the same between all pairs
of galaxies or clusters -- contrary to the simple vacuole model.
The expansion rate varies depending on the
matter and gravity field variations in the space
between the observer and the object. Therefore the really
measured Hubble parameter is rather
a stochastic variable that can be expressed relative to
the truly universal vacuum Hubble constant $H_V$ and the global
time-dependent average Hubble constant $H_0 = H_V(1+
\frac{\rho_M (t_0)}{\rho_V})$ . In any particular measurement
\bea
 H_i = H_V (1 + \frac{\rho_M (t_0)}{\rho_V} )^{1/2}(1 +
\Delta_i)\; \mathrm{km/s/Mpc}.
\eea
The deviation $\Delta_i$ must depend
on the spatial scale of measurement, related to the average matter
density contrast. 

\subsection{The local density contrast}

The situation
at small distances is also not the ideal one described by the
vacuole model. E.g., the conjunction
condition of Eq.16 may not be fulfilled in the Local Universe,
unless we consider a small volume around the Local Group. This is
because the mass of the Local Group is considerably less than the
mass of a uniform sphere with the cosmological density $\rho_M
(t_0)$ and the radius 5-10 Mpc. In fact, a cosmological volume
containing the non-relativistic mass $M$ of the Local group (dwarf
galaxies contribute negligibly within a few
percent accuracy) has the present-day radius
\be R (M) = [M /(\frac{4 \pi}{3} \rho_M)]^{1/3} = R_V (2\frac{\rho_V}{\rho_M})^{1/3}
 \simeq 2.2 \; \mathrm{Mpc}.
\ee As this is below typical distances between galaxy
groups locally, there may be a density deficit. This would not
influence the validity of the very local model that depends only
on the mass of the Local Group and the value of the vacuum
density, but it would naturally change the expected value of the
expansion rate of the Local Universe.

On the other hand, if we allow for an extended distribution of
dark matter beyond the 1 Mpc radius of the Local Group, perhaps
following a fractal density law (e.g. Teerikorpi et al. 1998;
Baryshev \& Teerikorpi 2002; 2005), then with the fractal
dimension $D = 1.2$  or the correlation exponent $\gamma = 1.8$,
the cosmological density would be reached around 3.2 Mpc. With $D
= 2$, this distance would be 8 Mpc. We do not know the exact
distribution of dark matter, but clearly it could add to the
total mass so that the conjunction condition is locally fulfilled.

Inspection of Fig.4 in Karachentsev et al. (2003b)
shows that within about 3 Mpc the Local Group is alone,
while within 6 Mpc there are six other
groups. Roughly, this would correspond to an average
matter density of $0.3 \rho_M$ within 3 Mpc, if there is no dark
matter related to the Local Group beyond 1 Mpc, while with the
dark matter distribution with $\gamma = 1.8$ the average matter
density would be close to $\rho_M$.
Hence, it seems that locally the average
matter density could be, roughly, in the range $(0.3 - 1) \rho_M$,
depending on the distribution of the dark matter around the
groups. This would
agree with other evidence that within a distance of 8 Mpc
there is only a small density excess, if any (Hudson 1992; Maccio et
al.2005). We note that generally a randomly placed observer
would be expected to be surrounded by a clear density excess
(c.f. sect.8 in Teerikorpi, Chernin, Baryshev 2005).

\subsection{The meaning of the local Hubble expansion rate}

Because of the relatively low matter content around us,
the domination of cosmic vacuum may be stronger in the Local Universe
than on the average over the Universe as a whole in the present
epoch. This special feature of the Local Universe could
offer better chances for vacuum measurements there.

First, if there is a density deficit, the local Hubble constant
$\bar H$ in general is less than the global constant $H_0$, and
closer here to the universal Hubble constant $H_V$, at least on sufficiently
long scales : $H_V < \bar H_0 < H_0$.
Secondly, dwarf galaxies close to the Local Group may be sensitive
indicators of the vacuum, especially if the LG mass is concentrated
within a small volume (the local cosmology model; sects. 2,3). We
discuss this in the next section.

To illustrate these considerations, we take a look at Eqs.1-10
of the local cosmology. If one assumes that the local flow
contains mostly parabolic or nearly parabolic trajectories,
then the very local Hubble rate is
\be \bar H_0(r) = \frac{\dot r}{r} = H_V(1 + \frac{2GM A^2}{r^3})^{1/2} =
H_V (1 + 2 (\frac{R_V}{r^3})^3)^{1/2}. \ee

When $2GM A_V^2/r^3 = 2 (R_V/r)^3$ is small, the flow is close
to the asymptotic regime, and the deviation of $\bar H_0$ from
$H_V$ is less than, say, 6\% at distances $r \ga 2.5 R_V \simeq 3-4$
Mpc.

Of course, there are distortions of the local
flow caused by the Virgo cluster and other environmental effects, but
these
may still be relatively weak at $r < 3-4$ Mpc.
This is supported by various
observational studies of the local Hubble flow, as summarized in
Teerikorpi, Chernin, Baryshev (2005). In these studies, the local
Hubble constant is found within the interval $56 - 70$ km
s$^{-1}$Mpc$^{-1}$, so that the central value here is quite close
indeed to the theoretically expected value of $H_V$.

\section{The shortest-distance indicators of dark energy}

The theory of the local flow (Secs.2,3) leads to two clear
predictions that may be falsified by observations:

(1) There is a lower limit for the flow velocities outside the
zero-gravity sphere. Indeed, the escaped galaxies have the energy
$\bar E \ge U_{max} = -\frac{3}{2} GM/R_V$, as is seen from Eq.6.
For any distance $r$, this inequality implies that the flow
velocity must be large enough:
\be \dot r(r) \ge V_0(r) \equiv H_V R_V (2 R_V/r + r^2/A^2
-3)^{1/2}.\ee

(2) There is a lower limit for the distances in the local Hubble
flow: $r \ge R_V$.

Prediction (1) is critical: if only a single
isolated and accurately observed galaxy is found to violate this
limitation, the theory or the values of its parameters must be
rejected or modified. Prediction (2) is
closely related to the observed kinematical structure of the flow
and may serve for effective measurements of dark energy density.

\subsection{Comparison with observational data}

Here we concentrate on the nearest
distances which are most sensitive to the test.
It is also here where the distances (TRGB, Cepheids) are expected to be
less influenced by selection effects causing systematic errors (Teerikorpi
1997; Paturel \& Teerikorpi 2005).

Indeed, recent data on the Local group vicinity (1-3 Mpc
from the group barycenter) are given by Karachentsev et al.
(2002; 2004). In the critical distance interval 1-2 Mpc, 8 galaxies
were studied. Their distances are measured with a typical
reported accuracy of 10\% and the accuracy of the velocities is
1-2 km/s. The galaxy SagDIG is the nearest to the Local group
barycenter: its distance is $1.2$ Mpc. Its velocity, 23 km/s
relative to the group barycenter, is the lowest in the flow. The
next closest galaxy is SexB with the distance of 1.6 Mpc and the
velocity of 111 km/s. The three galaxies, Antlia, NGC3109 and
SexA, have almost equal distances near 1.7 Mpc and the velocities
of 66, 110 and 94 km/s, respectively. The galaxy KKR25 is of the
distance 1.8 Mpc and the velocity 68 km/s. The galaxies E294-010
and KKH98 are located at the distance near 2 Mpc and have the
velocities 81 and 152 km/s.
We also add NGC300, with its Cepheid
distance (1.93 Mpc) recently measured by Gieren et al. (2004),
and the velocity 125 km/s and seven other galaxies with $r \leq 2.5$
Mpc from the
catalogue of Karachentsev et al. (2004). Thus we have all the galaxies
with known distances (from the TRGB or Cepheid method) between
1 Mpc and 2.5 Mpc.

\begin{figure}
\epsfig{file=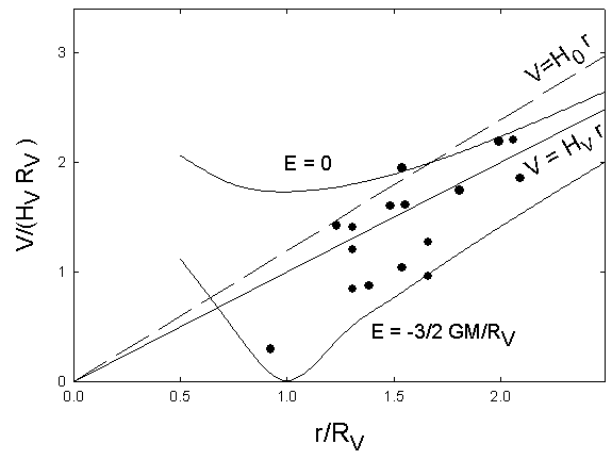,angle=0,width=9cm}
\caption{The Hubble diagram for very nearby galaxies,
with the units $R_V = 1.3$ Mpc and $H_V R_V = 78$ km/s have been
used for x- and y- axis, respectively. The linear
velocity-distance relation is shown by two straight lines,
one for the universal Hubble constant $H_V$, and the other for the
global Hubble constant $H_0$. The trajectories for the parabolic
motion ($E =0$) and for the lower limit velocity $V_0$ are given.}
\end{figure}

All the 16 galaxies with their barycentric distances from the LG
are plotted on the Hubble diagram  in Fig.1,
where the units $R_V = 1.3$ Mpc and $H_V R_V = 78$ Mpc have been
used for x- and y- axis, respectively. The linear
velocity-distance relation is shown by the two straight lines,
one for the universal Hubble constant $H_V$, and the other for the
global Hubble constant $H_0$. The trajectories for the parabolic
motion ($E =0$) and the lower limit velocity $V_0$ are also given.
We may see in Fig.1 that all the galaxies here obey the
restriction of Eq.28.

We may also see that -- in agreement with the discussion of
Secs.2,3 -- the two curves reveal a trend for converging to the
universal (with $H_V$) linear velocity-distance relation. The
lower limit velocity approaches the relation rather closely at
the distances 5-7 Mpc (mentioned also in the section above) that
could provide another sensitive interval for an observational test.

\subsection{Towards weighing dark energy}

In global cosmology, the dark energy of cosmic vacuum has
revealed itself in observation (Riess et al. 1988, Perlmutter et
al. 1999) at very large distances comparable with the vacuum
length $A$ or the horizon radius. The corresponding redshifts are
near
\be 1 + z_V = [2 \rho_V/\rho_M (t_0)]^{1/3} \simeq 1.7. \ee At $z
= z_V \sim 0.7$, the global cosmological expansion transits from
deceleration to acceleration.

The redshift $z_V$ is a clear global counterpart of the local
zero-gravity radius $R_V$: the local flow transits from
deceleration to acceleration at $r = R_V$. The difference,
however, between them is essential. In the global cosmology, the
balance of gravity and antigravity occurs only for one moment of
time (7-8 Gyr ago) and simultaneously for the whole Universe. On
the contrary, in the local cosmology, the balance exists for $\sim
10$ Gyr, but only at the zero-gravity surface. In a sense, what
is temporal in the global cosmology is spatial in the local one,
and vice versa.

Could observations at short distances near $R_V = 1-2$ Mpc  be as
useful and important as the observations at high redshifts near
$z_V \sim 1$?

The theory prediction (2) above indicates that the distance $R_V$
plays a key part in the dynamics of the local flow. As
first mentioned in Paper I, the local Hubble flow starts very
nearly at the zero-gravity sphere, if not exactly at $r = R_V$. This
is hardly just a funny numerical coincidence with no physical
sense. Rather, the coincidence is quite suggestive: it points out
the shortest distances where the dark energy may be detected and
measured
from its effect on velocities.

Indeed, as is seen from Fig.1 (and from the basic equations of the
theory as well), the velocities of  the flow trajectories have
minima at $r = R_V$. The location the minima is the same for all
the trajectories, independently of their initial conditions. If
the location of the velocity minima $r = R_V$ is found from
observations, the universal Hubble constant may be directly
determined (following from the definition of Eq.3):
\be H_V = 60 (\frac{M}{1.5 \times 10^{12} M_{\odot}})^{1/2}(\frac{1.3
Mpc}{R_V})^{3/2} \;\;\; \mathrm{km/s/Mpc}. \ee

Similarly, the density of the vacuum dark energy is
\be \rho_V = 7 \times 10^{-30} (\frac{M}{1.5 \times 10^{12}
M_{\odot}})(\frac{1.3 Mpc}{R_V})^{3}\;\;\; \mathrm{g/cm^3}. \ee

Now we turn to the data of Fig.1 again and inspect them for
the velocity minimum location.
Noting that
the velocities appear to be still decreasing around $r/1.3 \mathrm{Mpc}
\approx 1.4$ a  robust
conservative estimate follows for the upper limit value of the
zero-gravity radius:
\be R_V \la 1.8 \;\;\; Mpc. \ee This agrees
with the theoretically expected provisional value 1.3
Mpc that we had in mind above. Combined with Eqs.30,31, this
leads to observational restrictions on the values $H_V^2/M$ and
$\rho_V/M$. Assuming $M = 1.5 \times 10^{12} M_{\odot}$, one has
from Eq.32 an independent estimate for the lower limit of
the dark energy density in the Local Universe:
\be \rho_V \ga 3 \times 10^{-30} \;\;\; g/cm^3. \ee

If the lowest velocity in the flow (that of the galaxy SagDIG) is
tentatively identified as showing the location of the
velocity minimum in the flow, we may get $R_V = 1.2 \pm 0.1$ Mpc.
With this (not so reliable) estimate, the value spread is somewhat
narrower and the accuracy of the estimate is about 20 \%:
\be \rho_V = (9 \pm 2) \times 10^{-30} \;\;\; g/cm^3. \ee

We give these values as examples of what could be done with
higher precision using considerably more numerous distance
determinations in the interval 1 -- 2.5 Mpc.
One expects many more still unmeasured dwarf galaxies within
these distances (Karachentsev et al. 2004).
In any case, the very initial local flow with the shortest
distances and lowest redshifts -- in drastic contrast to the
important, but very difficult high redshift SN studies -- have proved
to be a sensitive indicator and `weighing device' 
of the dark energy.

\section{Discussion}

The current standard cosmological model may actually be seen as
containing two fundamental and equally important sectors. One is
the traditional global cosmology for the spatial scales of
300-1000 Mpc and more, while the other is the local cosmology.
The former studies on largest scales the uniform and evolving
space-time; the latter deals with the non-uniform static
space-times of the Local Universe and many (or infinite number)
of other more or less similar local regions. The global sector
uses the Friedmann solution as the
theory basis, while the local one needs a non-Friedmann theory
that we proposed and described above (Secs.2,3) in a simple
analytical version. We explained also (Sec.3) how the two sectors
of the whole model might be linked to each other, -- at least, in
our example of an ideal vacuole scheme.

The omnipresent and perfectly uniform dark energy of the cosmic
vacuum is the major physical factor which unifies the two sectors
of the model. It is especially important that, in the present-day
Universe, the vacuum with its antigravity dominates
dynamically on global scales and also within individual local
volumes outside the zero-gravity surfaces.
On the global scales, the vacuum antigravity accelerates the
general cosmological expansion produced by the initial Big Bang.
Locally, the vacuum antigravity makes dwarf galaxies ( mostly,
but also galaxy groups outside each others' zero-gravity spheres)
move from the Local Group and apart from each other with
acceleration, finally establishing the Hubble flow.

On both global and local scales, the vacuum provides the motions
with a common asymptotic behaviour that is the self-similar
parabolic expansion entirely controlled by the vacuum antigravity.
This dynamical regime is an attractor for a wide range of real motions
on the cosmic vacuum background, including the Friedmann
uniform expansion and the local spatially non-uniform flow.
In this limit, the expansion follows the exponential law, both
globally and locally: $r \propto R \propto exp(t/A_V)$. As a
result, the linear velocity-distance relation establishes, $\dot
r \propto r;  \dot R \propto R$, i.e. the Hubble expansion law.
Both locally and globally, the expansion rate is given by the
same universal physical constant related directly to the vacuum
density, in the limit: $ H_V = \dot r/r = \dot R/R = 1/A_V \simeq
60$  kms$^{-1}$Mpc$^{-1}$. Correspondingly, the maximally
symmetric space-time described by the de Sitter metric serves as
the universal asymptotic for both Friedmann global non-static
space-time and the non-Friedmann local static space-time.

The concordance data for global cosmology and the data for the
Local Universe indicate that the present-day states of both
global and local expansion flows are not far from the universal
asymptotic regime. It is because of this fact that the observed
local expansion rate is so close to the global Hubble constant
(which has been rightly considered as a big cosmic puzzle -- see
Sandage 1999). Thanks to the same fact, the velocity
dispersion is rather low in the local expansion
flow (which has also been a true mystery).

The simplest analytical version of the local model described in
this paper is supported by computer simulations (Chernin et al.
2004, 2005) which extend the model into the interior of the
zero-gravity surface. The simulations describe the Local Group as
a binary (the Milky Way and M31) galaxy, in accordance with the
Kahn-Woltjer (1959) classical model. They confirm that the
differences from the point-mass analytical model are indeed small
outside the zero-gravity surface. The binary produces relatively
small deviations from spherical symmetry in the local gravity
field at $r \ge R_V$, and also a  relatively
weak time dependence of the gravity field at these distances. In
particular, the simulations show that the zero-gravity surface is
almost perfectly spherical at present.

The simulations describe all the history of the local flow during
12-13 Gyr since the formation of the Local Group. They enable us
also to study the "initial conditions" for the local flow, and
indicate that the Little Bang model (Byrd et al.1994) for the
initial chaotic state of the Local Group (see also van den Bergh
2003) could produce a flow of galaxies across the zero-gravity
surface and their radial velocities are later amplified by the
antigravity push of the cosmic vacuum outside the zero-gravity
surface.

Our analytical theory and simulations are in good agreement with
the large N-body $\Lambda$CDM cosmological simulations (Ostriker
and Suto 1990; Suto et al. 1992; Governato et al. 1997; Nagamine
et al. 1999, 2000; Macci\`{o} et al. 2005).
These simulations give important
insights into the early history of the Local Universe when its
matter started to separate from the general cosmological
expansion at the cosmic age of 0.5-1.5 Gyr. The picture is
complementary to ours and demonstrates that the Local Universe
with a massive group in its center and the quiescent expansion outflow
is rather typical for scales of a few
megaparsecs. Such local structures emerge in a natural way from
the standard picture of gravitationally instability with the
Harrison-Zeldovich initial spectrum. Very importantly, these
`global' simulations have shown -- confirming the original
approach of our Paper I -- that the dark energy is necessary for
explaining cool local flows.

\section{Conclusions}

We summarize briefly the main results of the present study:

\begin{itemize}
\item[$\bullet$]{ A non-Friedmann model with a non-uniform static space-time
was proposed for the neighborhood of the Local Group in order to
describe the trajectories of galaxies flowing across the local
zero-gravity surface at $r=R_V \simeq 1.3$ Mpc. The model is given
both within Newtonian theory and General Relativity.}
\item[$\bullet$]{It was shown that this local space-time, when treated within General
Relativity, may be embedded in the global Friedmann space-time,
and the Local Universe may described as a spherical vacuole,
according to the Einstein-Straus solution.}
\item[$\bullet$]{We emphasize the difference between the truly universal
Hubble constant $H_V$, depending just on the constant vacuum
density, and the measured local or global Hubble constant $H_0$
that depend both on spatial matter distribution and scale
and on the cosmic time.}
\item[$\bullet$]{We have inspected the predicted trajectories of the
out-flowing galaxies
and demonstrated analytically the efficiency of vacuum cooling,
which is higher than the usual adiabatic cooling: all
trajectories within the vacuole tend towards the Hubble relation
with the universal vacuum Hubble constant $H_V$. This gives a
natural explanation for both the low velocity dispersion in the
flow and its closeness to the global Hubble constant.}
\item[$\bullet$]{Still another problem, the absence of turning-down of the
velocity-distance relation close to the Local Group (Ekholm et al.
2001; Thim et al. 2003) and in general the absence of the infall towards
local groups (Whiting 2005) may be explained by the present
model, where the out-flowing dwarfs remain close to the universal
Hubble relation and have a positive minimum velocity at $r = R_V$.}
\item[$\bullet$]{An interesting result is that all out-flowing trajectories have
a velocity lower limit at any distance, corresponding to the total
energy just required to reach $r=R_V$}
\item[$\bullet$]{Preliminary comparison with available distance and
velocity data for  nearby galaxies,
1 - 2 Mpc, shows agreement of the predicted trajectories with $E
\leq 0$ and may suggest the expected minimum around 1.3 Mpc. The
local Hubble diagram $V/(H_VR_V)$ vs. $r/R_V$ is a promising
instrument for such studies.}
\end{itemize}

Finally, the local Hubble flow is not so mysterious any more, but it may
rather be seen now as a natural tool for the detection and
study of the vacuum antigravity and the dark energy. In fact,
we have already used it for these
purposes since our Paper I. Combining our theory and computer simulations with
precision observations, we have discovered dark energy in the Local
Universe.
\begin{itemize}
\item[$\bullet$]{The flow provides strong evidence for the very fact of the
presence of dark energy in the Local Universe on the small scales
down to $\sim 1$ Mpc; the local DE density seems to be
close to that measured on the largest cosmological scales.}
\item[$\bullet$]{The dark
energy in the Local Universe can be well described in the
simplest way as the cosmological constant or cosmic vacuum with
the density-to-pressure ratio equal to -1; this agrees with the
concordance global cosmological model.}
\item[$\bullet$]{The dark energy with these properties might have existed
unchanged for 12-13 Gyr in the Local Universe, since the formation of the
Local Group; the global cosmology has no data on the dark energy
at such distances/times.}
\end{itemize}

These conclusions need and deserve further observational verification. We
advocate for extensive systematic high precision observations of
galaxies, focused on short distances, especially of 0.5-2 Mpc. The
advantages of such data are clear. 
These observations may give us increasingly exact information about
the dark energy. In particular, the local measurement of
the dark energy density can be much improved as compared
with our preliminary estimate here. Strong constraints may also be
obtained on (1) deviations from the standard density-to-pressure
ratio in the present-day Local Universe, (2) the time variation of
the dark matter density for a period of 12-13 Gyr since the
formation of the Local Group, and (3) the time variation of the
density-to-pressure ratio during the same long cosmological time.

Observations near the zero-gravity surface of the radius $R_V
\sim 1$ Mpc may prove to be as useful as the observations near
the redshift $z_V \sim 1$ of the zero-gravity moment in the
global Universe. With increasingly precise measurements of
distances and velocities in the Local Universe, the local sector
will be a central arena of further progress in cosmology.

\begin{acknowledgements} We appreciate valuable discussions with
Yury Efremov, Mauri Valtonen and our other colleagues at Sternberg
Institute and Tuorla Observatory.
\end{acknowledgements}

{}

\end{document}